\documentclass[pra,floatfix,preprint,nofootinbib,eqsecnum,12pt]{revtex4}
\usepackage{epsfig}
\usepackage{bm}
\usepackage{amsmath}
\usepackage{setspace}
\input epsf
\numberwithin{equation}{section}

\def\ed{ed.~by~}

\def\Ph{\hat P}
\def\Ch{\hat C}

\def\w0{w_0}
\def\bh{{\bar h}}
\def\be{\begin{equation}}
\def\ee{\end{equation}}
\def\cO{{\cal O}}
\def\Sh{{\hat S}}
\def\qh{{\hat q}}
\def\rhoh{{\hat \rho}}
\def\rhot{\tilde\rho}

\begin{document}
\vspace{1cm}

\title{  Alternative Decohering Histories in Quantum Mechanics\footnote{This is an early article in the authors' development of the decoherent  (or consistent) histories quantum mechanics of closed systems that is applicable to the universe as a whole.   It is of interest today as a compact statement of the program to explain, rather than posit,  the quasiclassical realm of every day experience from theories of the universe's quantum state and dynamics.  Ideas like fullness and a measure for classicality first appear here.  The article appeared in the Proceedings of the 25th International Conference on High Energy Physics, Singapore, August, 2-8, 1990, ed. by K.K. Phua and Y. Yamaguchi (South East Asia Theoretical Physics Association and Physical Society of Japan) distributed by World Scientific, Singapore (1990). It has been reset and posted to arXiv by  the junior author for better accessibility and as a historical record of research. Except for minor obvious corrections and for consistency with later terminology, no additions or modifications of the original text have been made nor have the references been updated.  A list of classic references to decoherent histories quantum mechanics is appended as well as a list  of all the authors' joint papers. \\}}


\vspace{1cm}

\author{Murray Gell-Mann}
\affiliation{California Institute of Technology, Pasadena, CA 91125}
\author{James  Hartle}\email{hartle@physics.ucsb.edu}
\affiliation{Department of Physics, University of California, Santa Barbara,  93106, USA}

\date{\today }

\begin{abstract}
We continue our efforts to understand, within the framework of the quantum mechanics of the universe as a whole, the quasiclassical realm of familiar experience as a feature emergent from the Hamiltonian of the elementary particles and the initial condition of the universe. Quantum mechanics assigns probabilities to exhaustive sets of alternative decoherent histories of the universe. We introduce and define the notion of strong decoherence. We replace the notion of maximal sets of alternative decohering histories by defining the more useful concept of ``full'' sets of alternative strongly decohering histories. These full sets fall  into equivalence classes each of which is characterized by a basis in Hilbert space. Finally we describe our continuing efforts to find measures of classicality --- measures that could be applied to such full sets  of alternative strongly decohering so as to characterize a quasiclassical realm.
\end{abstract}



\maketitle

\bibliographystyle{unsrt}
 

%

\section{Introduction}
\label{intro}
This contribution is a sequel to our {\it Quantum Mechanics in the Light of Quantum Cosmology} which has appeared in nearly identical forms in the proceedings of two conferences \cite{C1A, C2A}.
We are continuing our efforts to understand quantum mechanics and make it intelligible to others, not just as a way of predicting the statistical distribution of results in a sequence of reproducible laboratory experiments (such as proton-proton scattering at a given energy) but also as a framework for describing everything that goes on and has gone on in the universe.

As in our first communication, we shall attempt to characterize a ``quasiclassical realm" by utilizing such concepts as decoherence, maximality, and classicality.   Here we shall exhibit what we believe to be significant progress in defining and understanding decoherence, finding an improved version of maximality, and searching for the meaning of classicality. Our ultimate purpose remains, of course, to show how one or more quasiclassical realms emerge from quantum mechanics, elementary particle  dynamics and an initial condition for the universe, without the necessity of postulating the existence, outside of quantum mechanics, of a "classical realm," with which contact is made in a "measurement situation''. 

We can describe our previous work in terms of a certain way of looking at quantum mechanics, based on the following elements \cite{C3}:
\begin{enumerate}
\item{}  The possible sets of alternative fine-grained histories of the universe, which are   the most refined descriptions allowed in the theory.

\item{} A notion of coarse graining by which the fine-grained histories are partitioned into exhaustive sets of mutually exclusive classes. Each such class is a coarse-grained history and each such set is a set of alternative coarse-grained histories of the universe. Further partitioning results in further coarse graining, so that there is a partial ordering of all the exhaustive sets of exclusive alternative histories.

\item A complex ``decoherence functional'' $D(h', h)$ of any two alternative histories  $h'$ and $h$ from a set (fine-grained or coarse-grained), with the properties   
\begin{enumerate}
\item{Hermiticity:}
$D(h,h')=D^{*}(h',h)$.\hfill (1.1). 
\item Positivity:
$D(h,h)\geq 0$. \hfill (1.2).
\item Normalization:   
$\sum_{(h,h')} D({h,h'})=1$. \hfill (1.3) 
\end{enumerate}
\setcounter{equation}{3}

\item A superposition principle (with respect to coarse graining) obeyed by the decoherence functional: If the set $\{\bar h\}$ is a coarser graining of a coarse-grained set \{$h$\}, then $D$ satisfies the equation
\be
\label{1.2}
D(\bh',\bh ) =\sum_{h \in \bh'} \sum_{h \in {\bh}} D{(h', h}) .
\ee

\item  A decoherence condition for a set of alternative histories that permits probabilities to be assigned to those histories: when the decoherence functional obeys an appropriate restriction, the quantities 
\be
p(h)\equiv D(h,h)
\ee
obey the rules of probability calculus, including the sum rules
\be
\label{1.3}
p(\bh)=\sum_{h \in \bh} p(h), 
\ee
and then thus be identified as the probabilities of the histories $h$. 
\end{enumerate}

The necessary and sufficient condition for decoherence \cite{C4} takes the form
\be
\label{1.4}
Re D(h',h)=0, \quad h' \ne h
\ee
in our formulation, but we shall also discuss other, more restrictive conditions, that are sufficient but not necessary.

In our previous work we have discussed situations in which the condition of decoherence is obeyed to a high accuracy, but not necessarily exactly, so that probabilities are defined only in a good approximation. However, we shall deal here with exact decoherence and probabilities that obey the rules of probability calculus exactly.

 In some contexts it is useful to broaden the definition of $D(h', h)$ by allowing $h'$ and $h$ to belong to {\it different sets} of alternative histories; conditions \eqref{1.3} and \eqref{1.4}  are retained.

Our approach, along with the characterization of the fine-grained histories and the specification of the decoherence functional, can be used not only to describe familiar Hamiltonian quantum mechanics, including quantum cosmology, but even, should  that prove necessary, to generalize it slightly [3]. In this communication, however, as   in Ref. [1] and [2], we shall stick to ordinary quantum mechanics and also restrict the discussion still further by ignoring, for simplicity, two important complications introduced into quantum mechanics by gravitation:

\begin{enumerate}

\item	possible major quantum fluctuations of the spacetime metric, which make it difficult to define a background spacetime and a quantum-mechanical time $t$, along with a Hamiltonian $H$ for all the elementary particles and their interactions;  
\item	possible sums over topologies of a Euclideanized spacetime, with associated   "wormhole" or "baby universe" effects.
\end{enumerate} 

In our simplified treatment, then, the fixed background spacetime provides a well defined notion of time, so that the normal apparatus of Hilbert space, states, operators, and unitary evolution may be employed in defining the elements of quantum   mechanics discussed above, The basic  laws of phvsics are represented by $H$. the Hamiltonian, and by $\rho$ , the density matrix of the universe in the Heisenberg picture. Here $H$ describes the unified quantum field theory of all elementary particles and their interactions (for which superstring theory provides the only known candidate). The density matrix $\rho$ corresponds to the boundary condition near the beginning of the expansion of the universe; a possible candidate is given by the ``no boundary" proposal, in which $\rho$  is pure and determined by the same action function as the theory of the elementary particles. Local Heisenberg field operators have a dependence on time of the form 
\be
\cO(\vec x, t)=e^{iHt}\cO(\vec x, 0)e^{-iHt} .
\ee

Since we are ignoring some of the complications of quantum gravity, the universe is being treated, more or less, as a box with a fundamentally simple initial condition at a time $t_0$, when the Heisenberg and Schr\"odinger pictures are taken to coincide.

 As in our earlier work, we are discussing a set of alternative fine-grained or coarse-grained histories of the universe described by sequences of projection operators onto   ranges of values of Heisenberg operators at a succession of times $t_1, .., t_n$ with $t_0 < t_1 < t_2 < < t_{nÑ1} < t_n$. We have, at each time $t_k$, an exhaustive set of mutually   exclusive projection operators $P^k_{\alpha_k} (t_k)$, by which we mean that
 \be
 \sum_{\alpha_k} P^k_{\alpha_k} (t_k) =1 \quad \text{and}  \nonumber
 \ee
 \be
P^k_{\alpha_k}(t_k)P^k_{\alpha'_k}(t_k)=P^k_{\alpha_k}(t_k)\delta_{\alpha_k,\alpha'_k} .
\ee
Here, $k$ denotes the set of alternatives at a given time and $\alpha_k$ the particular alternative in the set. An individual history in the set corresponds to a particular sequence of projections, that is to a particular sequence of $\alpha's$. We shall sometimes use the  notation  $\alpha$ for a sequence $(\alpha_1,\alpha_2, \cdots,\alpha_n)$ and $C_\alpha$  for the product   $P^n_{\alpha_n}(t_n) \cdots  P^1_{\alpha_1}(t_1)$ of all the projection operators in the chain corresponding to an individual history.

The completely fine-grained sets of alternative histories correspond in this case to sequences of one-dimensional projection operators onto complete sets of states, with one such set of projection operators at each and every time. There are clearly many such fine-grained sets. These fine-grained histories may be partitioned by the following operation of coarse graining: A set of histories is a coarse graining of a finer  set if every projection defining the coarser-grained set is a sum of projections in the finer-grained set. Conversely the finer set is a fine-graining of the coarser one. Of course, two given sets need not be either fine or coarse grainings each other. Thus the operations of coarse and fine-graining define a partial ordering  of the set of all exhaustive sets of exclusive histories. The process of coarse graining terminates in the trivial case of a single $C$, which is the identity operator.

The possible sets of histories considered thus correspond to sequences of all possible sets of exhaustive and exclusive projections at all possible ordered sequences of times $ t_0 \leq t_1 \leq  t_2 < \cdots \leq t_n$. The decoherence functional for such coarse-grained sets is
\be
D(\alpha',\alpha) = Tr(C_{\alpha'}\rho C_\alpha^\dagger)
\ee
It is easy to see that it obeys the conditions (1.1)-(1.4).   The probability formula  (1.5) and the necessary and sufficient condition (1.7) for defining probabilities may  be combined in the fundamental formula
\be
\label{weak}
Re D(\alpha',\alpha)=p(\alpha) \delta_{\alpha',\alpha}
\ee
which we shall call the {\it weak decoherence condition} for the set of alternative histories involved.

In most familiar situations in which weak decoherence occurs, it is not only the real part of the decoherence functional $D(\alpha',\alpha)$ that vanishes for $\alpha' \neq \alpha$ but rather the whole quantity so that we have 
\be
\label{medium}
D(\alpha',\alpha)=p(\alpha) \delta_{\alpha',\alpha}
\ee
which we call the {\it medium decoherence condition} on sets of histories. It is the condition that we used in our earlier work.

With these concrete representations of the general elements described above, the description of our simplified formulation of quantum mechanics is complete.  

\section{Strong Decoherence}
\label{strong}
Let us treat first, for simplicity, the case of a pure density matrix
\be
\rho=|\Psi\rangle\langle\Psi|
\ee
which was discussed to some extent in Refs \cite{C1A} and \cite{C2A}. The decoherence functional $D(\alpha',\alpha)$ is then given by a scalar product
\be
D(\alpha',\alpha) = Tr(C_{\alpha'}\rho C_\alpha^\dagger) =  \langle \Psi | C_\alpha^\dagger C_{\alpha'} |\Psi\rangle
\ee
and we see that for the case of medium decoherence, the states $C_\alpha |\Psi\rangle$ that are non-vanishing are all orthogonal, with their norms giving their probabilities.  Since, 
the ``branches"  $C_\alpha|\Psi\rangle \neq 0$ are a set of orthogonal vectors in Hilbert space, there is obviously a set of exhaustive and mutually exclusive projection operators $R_\alpha$  that define orthogonal subspaces in which those vectors lie:
\be
\label{2-3}
C_\alpha |\Psi\rangle = R_\alpha |\Psi\rangle .
\ee 
where the normalization follows from the fact that $\sum_\alpha C_\alpha$ is the unit operator. In most situations the non-vanishing $C_\alpha$ do not form a complete set of states and therefore the projection operators $R_\alpha$ can be chosen in many different ways.

We can think of the operators $R_\alpha$  as representing generalized records of the histories $\alpha$, especially if they are expressed (as is always possible) as projection operators onto ranges of values of Heisenberg operators at a time $T$ such that $T \ge t_n >\ge. . . \ge t_1$ . As we mentioned in our previous work, the reconstruction of past history is most generally viewed as the assignment of probabilities to alternatives in the past, given present data as well as the initial condition and the dynamics. When there is a perfect correlation between each coarse-grained history in a set and particular values for certain present data, we may say that the data constitute generalized records of the  past. (We use the term "generalized" because a record in a stricter sense would be defined to exhibit persistence over an appropriate time scale as well as correlation with the past.) We can say that an individual coarse-grained history ``happened'' if it is fully correlated with our present records, which are the only means of discriminating among the various possible histories of the past permitted by the initial condition and the dynamics.

We see that for a pure $\rho$ the condition \eqref{medium}  for medium decoherence implies, through \eqref{2-3}, the condition
\be
\label{strng}
C_\alpha\rho=R_\alpha \rho,   \quad   (\text{with} \  R_{\alpha'}R_\alpha = R_\alpha \delta_{\alpha'\alpha} ), 	
\ee
which we shall call {\it strong decoherence} and which we have identified with the existence of generalized records.
  The name ``strong decoherence" is justified because that condition is no longer necessarily implied by medium decoherence when $\rho$  is impure. For example, consider
the density matrix corresponding to complete indeterminacy of the initial state:
\be
\rho_{\text{ind}}= 1/Tr(1).
\ee
  For this density matrix, the strong decoherence condition \eqref{strng}  would imply an operator identity between each $C_\alpha$ and the corresponding $R_\alpha$; but that can happen only in the trivial case where all the $P$ 's at the different times commute.
  
It is easy to show that strong decoherence implies medium decoherence, which implies weak decoherence.
  
In conventional discussions of ``measurement theory," one often considers two non-commuting quantities being measured by two other quantities (positions of dials  etc.) that do commute, so that
\be
P^2_{\alpha_2}(t_2)P^1_{\alpha_1}(t_1)\rho=Q^2_{\alpha_2}Q^1_{\alpha_1}\rho
\ee
where 
\be
[P^2_{\alpha_2}(t_2), P^1_{\alpha_1}(t_1)] \neq 0
\ee
but 
\be
[Q^2_{\alpha_2}, Q^1_{\alpha_1}] =0
\ee
We can then regard $Q_{\alpha_2}Q_{\alpha_1}$ as a single projection operator $R_{\alpha_2\alpha_1}$, and we have a simple but familiar example of condition \eqref{strng}.

\section{Full Sets of Decohering Histories}
In conventional discussions of ``measurement theory," the chains would be composed of projection operators onto ranges of values of operators ``measured" by some particular ``observer" or set of communicating ``observers." We would like, however,   to escape, as much as possible, from an observer-centered description of quantum mechanics. One way of doing that, which we treated earlier \cite{C1A,C2A} , is to discuss maximal sets of coarse-grained decohering histories, that is, sets that are maximally fine-grained consistent with decoherence. (Recall that in this paper, in contrast to previous work, we are treating just the special case of perfect decoherence.)

Any set of coarse-grained histories available to an ``observer" (or composite ``observer is evidently a coarse graining of many different maximal sets of alternative histories and the amount of additional coarse graining is, of course, enormous in each case. Conversely, any of those maximal sets can be obtained by fine-graining the original set of histories.

By using maximal sets of alternative coarse-grained histories obeying one of the decoherence conditions \eqref{weak}, \eqref{medium}, \eqref{strng},  we are not tied to measurements available to any particular ``observer" or composite ``observer," and we can include projections onto alternative ranges of values of operators referring to places and times for which no sort of ``observer" was available.

In our previous work, we remarked that truly maximal sets of alternative decoherent histories often involved the presence of a great many redundant projection operators that merely repeated in new disguises the same information contained in other projection operators. Therefore we introduce here an alternative concept that captures much of what was desirable in the notion of maximality without requiring vast amounts of redundancy.

We recall that a set of alternative strongly decohering histories is characterized by a set of generalized record operators $R_\alpha$, mutually orthogonal projection operators such that  
\be
C_\alpha\rho=R_\alpha \rho .
\ee
  We noted that, for a pure $\rho=|\Psi\langle\rangle\Psi|$, the non-vanishing $C_\alpha|\Psi\rangle$ are not necessarily a complete set of states and therefore the $R_\alpha$ are not unique. However, we can
keep fine-graining the chains of projection $C_\alpha$, maintaining strong decoherence, to the point where the non-vanishing members of the set of mutually orthogonal states $C_\alpha|\Psi\rangle$ do form a complete set of states, a basis in Hilbert space, and the set of $R_\alpha$ becomes a unique complete set of one-dimensional projections. A similar construction is possible in the impure case: the chains $C_\alpha$ can be fine-grained, maintaining  strong decoherence, until the projections $R_\alpha$ form a unique and complete set of one dimensional projections in Hilbert space, thus defining a basis. We will describe this situation as one in which the alternative strongly decohering histories form a {\it full} set.

There are, of course, many different sets of alternative strongly decohering histories with the same set of generalized records $R_\alpha$, and that is still true for full sets of histories. There are thus equivalence classes of full sets of histories defined by their common basis in Hilbert space or, what is the same thing, by the complete set of one-dimensional projection operators $R_\alpha$ onto those basis states.

By considering equivalence classes of full sets of histories, we achieve the same kind of independence of specific ``observers" that we achieved before by considering maximal sets of histories.

Now let us take a given equivalence class of chains $C_\alpha$ , corresponding to a particular basis in Hilbert space and to a complete set of one-dimensional projections $R_\alpha$. We can, if we like, express those projection operators as projections onto ranges of values of some Heisenberg operators at any time, for example a time $T$ such that
 $T\ge t_n, \ge \cdots \ge t_1$ and regard them as generalized records. We can also express them in terms of Heisenberg operators at any other specific time, for example $t_1$. Thus   one full set of histories in the equivalence class consists of completely fine-grained projections at any single time.
 
Within the same equivalence class, we can make use of a kind of redundancy described in our earlier work to exhibit a full set of histories that is also maximal  as well as completely fine-grained. The same operators $R_\alpha$ are expressed in terms of Heisenberg operators at many times and strung together in chains to make the $C_\alpha$. When all times (say between $t_0$ and $T$) are represented, we have a completely fine-grained maximal set of histories over the time interval, but constructed in a   trivial manner. In a similar fashion, given a set of histories described by chains $C_\alpha$, it is always possible to refine the chains so that there are projections at every time
by mindlessly interpolating projection operators identical to those already there but expressed in terms of Heisenberg operators at intermediate times.

For a pure $\rho =|\Psi\langle\rangle\Psi|$  it is easy to show that there are other types of trivial redundancy, making use of the state vector $|\Psi\rangle$, that allow the construction of other   completely fine-grained full sets of histories in the given equivalence class. Consider a set of exactly decohering histories specified by chains  $C_\alpha =P^n_{\alpha_n} (t_n) \cdots P^1_{\alpha_1}(t_1)$.
Successive projections in this chain define successive resolutions of the initial state vector $|\Psi\rangle$ into orthogonal vectors. At time $t_k,$ for example, these vectors are 
\be
\label{resolved}
P^n_{\alpha_k}(t_k) \cdots P^1_{\alpha_1}(t_1) |\Psi\rangle .
\ee
For a full set of histories, these successive resolutions terminate in a complete set of orthogonal vectors.  At each and every time $t$ between $t_0$ and $T,$ it is possible to interpolate in the chain of $P$'s one-dimensional projections onto any complete set of orthogonal states that includes the resolved set of vectors \eqref{resolved}, without affecting decoherence, fullness, or membership in the equivalence class. Thus every set of histories in the equivalence class is completely fine-grained, albeit in a trivial way.

We should also recall that, from a sequence of sets of projections in a given equivalence class, it is possible to construct another sequence simply by reassigning the times, leaving the order of times alone. Furthermore, given one equivalence class, it is possible to construct another by means of a unitary transformation that preserves $\rho$. Such constructions, as well as the types of redundancy described above, show that there are many full sets of histories that are stretched out in time in uninteresting  ways.
  
Now what kinds of sets of alternative histories {\it do} interest us in our effort to  describe a quasiclassical realm? We do not expect complete fine-graining at any  individual time. Rather we expect fullness to be achieved by sets of projections at a sequence of times, where the projection operators at different times are not merely  the same operators repeated over and over under different names, but  different sets   of operators, at the various times, that gradually determine, through the chains $C_\alpha$, a full set of histories. In Section \ref{towards}; we discuss some measures that may help us to   define classicality for a full set of histories, whether within an equivalence class or over all equivalence classes.

\section{Qualitative Features of a Quasiclassical Realm} 

In Refs.\cite{C1A} and \cite{C2A}  we discussed operators corresponding to hydrodynamic variables, that is, averages over suitable volumes of space of densities of exactly or approximately conserved quantities. The volumes are chosen small enough so that the matter enclosed is roughly in equilibrium and large enough so that the matter has sufficient inertia to resist the buffeting of most quantum fluctuations and classical statistical fluctuations. The choice of density operators and of volumes can vary with space and time and also effectively with the ``branch" in the branching histories of the universe, as described at greater length in Refs. \cite{C1A,C2A}. With suitable choices of the hydrodynamic variables, we expect there to be projection operators $P^n_{\alpha_n}, \cdots , P^1_{\alpha_1} $ onto ranges of values of those operators at a succession of times $t_1, \cdots, , t_n$, such that the chains $C_\alpha = P^n_{\alpha_n}, \cdots , P^1_{\alpha_1}$ strongly decohere (at least to an excellent approximation)  with respect to the density matrix $\rho$. Furthermore, we expect that the $P$ 's behave  roughly as if the corresponding hydrodynamic variables obeyed a closed set of classical (deterministic) equations of motion. These are not the fundamental equations of motion of the elementary particles, but rather a set of phenomenological equations including the effects of dissipation, etc.

Of course, occasional large fluctuations will disturb the operation of these equations, creating the phenomenon of branching or fanning out of orbits. In addition, as we have already remarked, even the choice of variables may be altered by large fluctuations. It is just because of such branch dependence that we must deal with alternative histories of the universe rather than with the fate of a predetermined set of variables.

Now consider the small fluctuations, which are largely resisted by the inertia of the hydrodynamic variables as they follow their phenomenological equations of motion, in between perturbations that cause major branchings. Those small fluctuations still play a crucial role in effecting decoherence of projections onto ranges of  values of   hydrodynamic variables, by carrying off appropriate quantum phases.

An instructive example, used by Joos and Zeh \cite{C5}, is provided by the photons of the $3^o$ K radiation encountered by a planet, or even a sizable dust grain, as it moves through the solar system Ñ the classical motion of the center of mass is not much affected, but successive positions at short intervals of time are made to decohere, at least approximately. The struck photons, moving off toward infinity, carry with them information about the collision; if we imagine the photons continuing undisturbed, then projections into their states of motion will become parts of the generalized records $R_\alpha$  at the ends of the histories. The correlation between different, orthogonal states of the photons and different past positions of the planet or dust grain produces    the decoherence of successive positions of that object.
It is the widespread occurrence of such mechanisms in the universe that ensures the habitual decoherence of suitably defined hydrodynamic variables. We call such habitually decohering variables quasiclassical.

In Refs. \cite{C1A} and \cite{C2A} we were dealing with the idea of a quasiclassical realm as a maximal set of alternative decohering histories further characterized by  some suitable measure of classicality (which we called ``classicity"). Now we propose to consider instead full sets of histories and define a quasiclassical realm by some process of optimization over those. With this new approach, a quasiclassical realm will not be a maximal set of alternative decohering histories, with all the redundancy that that implies. Thus, for example, it might turn out that a full set of histories constituting a quasiclassical realm could be characterized just by projections onto  ranges of values of suitable hydrodynamic variables at suitable instants of time. Such a full set of alternative strongly decohering histories would certainly not be maximal.

Finally, let us recall the idea of a ``measurement situation" \cite{C1A,C2A} (independent of whether or not there is any ``observer" --- i.e., an IGUS = information gathering   and utilizing system ---  present to make a ``measurement"). Our basic idea is that a measurement situation arises when the projection operator referring to some variable   is fully correlated with some of the projection operators of a quasiclassical realm. If the projection operator corresponding to the ``measured quantity" is not already present among the projection operators of the quasiclassical realm, then adjoining  it is a fine graining, and one that preserves decoherence. In our earlier work, where we treated the quasiclassical realm  as a maximal set of histories, the projections referring to the ``measured quantity" had to be included among the variables in that maximal set. Now, with the quasiclassical realm as a full but not maximal set of histories, the projections referring to the "measured quantity" will in many cases not be included among the projections of the quasiclassical realm. For example, if
projections referring to the quasiclassical hydrodynamic variables are sufficient to give the quasiclassical realm, then measured quantum variables will not be included, but would instead yield projection operators that could be adjoined to give a finer-grained set of histories in the same equivalence class.


\section{Towards Classicality}
\label{towards}
We should like to inquire whether the characterization of a quasiclassical realm  could be made rather precise and quantitative, and also stated in such a form that the quasiclassical (say hydrodynamic) variables would emerge as a consequence of $\rho$  and $H$ and not have to be put in as an assumption.

The simplest possibility we can envisage for describing the most classical sets of alternative decohering histories is to try to find a suitable quantity to maximize over the full sets of alternative, strongly decohering, histories constituting an equivalence class characterized by a basis in Hilbert space, and then try to find some other quantity 
to maximize over those equivalence classes. (If the two quantities should turn out   to be the same, then we could dispense, for the optimization procedure, with the  equivalence classes.)

Of course it may easily turn out that something more complicated than such a one- or two-step procedure is required.

Somewhere in this process, we must escape from the purely algebraic manipulations in Hilbert space that we have discussed so far and bring in the dynamics. As mentioned in Refs. \cite{C1A}  and \cite{C2A}, the definition of classicality must refer to the Hamiltonian, expressed in terms of fields, as well as the density matrix $\rho$. Otherwise (to take the case of a pure $\rho$  an example), there would be no physical distinction between one state vector $|\Psi\rangle$  in Hilbert space and another. In particular, if the alternative histories of the quasiclassical realm are correspond, at least in part, to the variation with time of certain quasiclassical variables, then there must, in many cases, be some degree of relationship between \{$P^k_{\alpha_k} t_k)$\} and  
\{$\exp{[iH(t_k-t_{k-1})}]$ 
$P^{k-1}_{\alpha_{k-1}}{(t_{k-1})} $
$\exp{[-iH(t_k-t_{k-1})]}\}$.

Let us consider an example of a quantity that measures, to some extent, the strength of this relationship among the $P$'s at different times. For each set of Heisenberg projection operators $\{P^k_{\alpha_k}\}$ at a given time $t_k$, we construct the corresponding  Schr\"odinger  projection operator 
\be         
\Ph^k_{\alpha_k} \equiv \exp{[-iH(t_k-t_{0})}]
P^{k}_{\alpha_{k}}{(t_{k})} 
\exp{[iH(t_k-t_0)]}.
\ee
We also define the Sch\"odinger chain
\be
\label{chain}
{\hat C}_\alpha\equiv \Ph^n_{\alpha_n} \Ph^{n-1}_{\alpha_{n-1}} \cdots \Ph^2_{\alpha_2} \Ph^1_{\alpha_1} .
\ee 

 
As in (2.5), we imagine that the dimension $Tr({1}) $ of Hilbert space is finite and utilize $\rho_{ind}\equiv  1/Tr(1) $. We can then define formal probabilities  
\be
\qh^\alpha \equiv Tr(\Ch \rho_{ind} \Ch_\alpha^{\dagger}) .
\ee
and consider the information content
\be
\label{infocontent}
\Sh \equiv -\sum_\alpha \qh_\alpha log \qh_\alpha .
\ee
of those formal probabilities.

We see that $\Sh$ tends to be low when the successive sets of Heisenberg projections  $P^k_{\alpha_k}(t_k)$ come close to being time translations of each other, so that the corresponding Schr\"odinger projections $\Ph^k_{\alpha_k}(t_k)$ come close to being identical. The quantity $\Sh$ thus reflects the departure from time correlation of the Heisenberg projection operators themselves, without any input from the actual density matrix $\rho$  of the universe.

The quantity $\Sh$ has some other interesting properties. It tends to increase upon the introduction into the sequence of projections $P^k_{\alpha_k}(t_k)$ of either of the two types of trivial refinement discussed in Section IV. Furthermore, it is not invariant under reassignment of the time $t_k$ or under unitary transformations of the Hilbert space that are not symmetries of the Hamiltonian. Thus a low value of $\Sh$  favors certain features that we associate with a quasiclassical realm.

Of course $\Sh$ by itself has no chance of characterizing such a realm, because it does not exploit the correlations present in the actual density matrix $\rho$, nor does it favor sets of histories in which the information contained in the $R_\alpha$ is stretched out in
time over a great many sets of projections $P^k_{\alpha_k}(t_k)$ . Furthermore, a low value of $\Sh$ fails to favor a high level of determinacy in histories, that is to say probabilities that are often near unity for successive projections correlated by classical phenomenological laws and often near zero for successive projections that are not so correlated.

A quantity that has some of these desired properties $\Sh$ not possessed by $S$ is the   information measure $S(\rhot)$ introduced in \cite{C1A}  and \cite{C2A}. We recall that in the definition of $S(\rhoh)$ we impose the conditions 
\be
Tr(\Ch_{\alpha'} \rhot \Ch_{\alpha}^{\dagger})=Tr(\Ch_{\alpha' }\rho \Ch_{\alpha}^{\dagger}) ,
\ee
and then maximize
\be
 S(\rhot)= -Tr(\rhot log \rhot)
\ee\
 subject to those conditions.

A low value of $S(\rhot)$ favors stretching of the histories and also favors probabilities for the histories that are as close to $0$ and $1$ as the branching allows.

On the debit side, the measure $S(\rhot)$ is indifferent to those redundant refinements of histories discussed in Section IV that are independent of $\rho$ and actually favors those   redundant refinements that are connected with $\rho$. Also, $S(\rhot)$  does not discriminate among sets of histories that are related by reassignment of the times $t_k$ or by unitary transformations of Hilbert space that preserve $\rho$.

 The examples of $\Sh$ and $S(\rhot)$ serve to suggest how, by the judicious use of more 
than one measure, we might be able to characterize classicality. Unfortunately, we have not so far progressed beyond hints as to how that could be done.

\acknowledgements
The work of MGM wass supported in part by the U.S. Department of Energy under contract No. DEAC-03081ER40050, and by the Alfred P. Sloan Foundation. The work of JH was supported in part by. NSF Grant PHY90-08502.

\eject

{\singlespace A list of joint papers by the two authors is appended so an interested reader can see the further development of these ideas.
The numbers refer to the junior author's publication list only as a ready to hand label. For instance further development of characterizations of classicality can be found in 97, 103, 116, 138, and 155. }
\vskip .2in

\centerline{PAPERS by MURRAY GELL-MANN and JAMES HARTLE}
\singlespace

\begin{itemize}

\item{87.} Quantum Mechanics in the Light of Quantum Cosmology, essentially identical versions of this paper appeared in {\sl
Complexity, Entropy, and the Physics of Information, Santa Fe Institute
Studies in the Sciences of Complexity VIII}, \ed W.H. Zurek, Addison-Wesley,
Reading, MA (1990) and in {\sl Proceedings of the 3rd International Symposium on
the Foundations of Quantum Mechanics in the Light of New Technology}, \ed S.
Kobayashi, H. Ezawa, Y. Murayama, and S. Nomura, Physical Society of Japan,
Tokyo, (1990), arXiv:1803.04605.

\item{94.}  Alternative Decohering Histories in Quantum Mechanics  
in the {\sl Proceedings of
the 25th International Conference on High Energy Physics, Singapore,
August,
2-8, 1990}, 
\ed K.K.~Phua and Y.~Yamaguchi (South East Asia Theoretical 
Physics Association
and Physical Society of Japan) distributed by World Scientific,
Singapore (1990).

\item{96.} Time Symmetry and Asymmetry in Quantum Mechanics and Quantum
Cosmology,
 in the {\sl Proceedings of the 1st International
Sakharov Conference on Physics}, Moscow, USSR, May 27--31, 1991, \ed
L.V.~Keldysh and V.Ya.~Fainberg, Nova
Science Publishers, New York (1992); and  in
{\sl Physical Origins of Time Asymmetry: Proceedings of the NATO
Workshop},
Magazon, Spain, September 30--October 4,
1991,
\ed J.~Halliwell, J.~Perez-Mercader, and W.~Zurek, Cambridge
University Press, Cambridge (1994) pp.~311-345; gr-qc/9304023.

\item{97.} Classical Equations for Quantum Systems, Phys. Rev., D47, 3345--3382, 1993; gr-qc/9210010.

\item{103.}Equivalent Sets of Histories and Multiple Quasiclassical
Domains; gr-qc/9404013.

\item{116.} Strong Decoherence  in the
{\sl Proceedings of the 4th (1994)
Drexel Conference on Quantum Non-Integrability:
Quantum-Classical Correspondence}, \ed D.-H.~Feng and B.-L.~Hu,
International Press of Boston/Hong Kong (1998);
gr-qc/9509054.

\item{138.}  Quasiclassical Coarse Graining and Thermodynamic Entropy, {\sl Phys. Rev. A}, {\bf 76}, 022104 (2007) , quant-ph/0609190.

\item{149.}  Decoherent Histories Quantum Mechanics with One `Real' Fine-Grained History, {\sl Phys. Rev. A} {\bf 85}, 062120 (2012); arXiv:1106.0767.

\item{155.}  Adaptive Coarse Graining, Environments, Strong Decoherence, and  \\ Quasiclassical Realms, {\sl Phys. Rev. A} {\bf 89}, 052125 (2014); arXiv:1312.7454,  DOI:  http://dx.doi.org/10.1103/PhysRevA.89.052125

\end{itemize}

\end{document}